\documentclass[preprint]{revtex4-1}
\usepackage{graphicx}
\usepackage{color} 
\begin{document}
\title{
The reversibility and first-order nature of liquid-liquid transition in a molecular liquid} 
\author{Mika Kobayashi}
\author{Hajime Tanaka} 
\email{tanaka@iis.u-tokyo.ac.jp}
\affiliation{Department of Fundamental Engineering, 
Institute of Industrial Science, {University of Tokyo, 4-6-1 Komaba, Meguro-ku, Tokyo 153-8505, Japan}}
\date{May 31, 2016}
\begin{abstract}
{\bf  Liquid-liquid transition is an intriguing phenomenon in which a liquid transforms into another liquid via the first-order transition. For molecular liquids, however, it always takes place in a supercooled liquid state metastable against crystallization, which has led to a number of serious debates concerning its origin: liquid-liquid transition vs. unusual nano-crystal formation. Thus, there have so far been no single example free from such debates. Here we show the first firm experimental evidence that the transition is truly liquid-liquid transition and not nano-crystallization for a molecular liquid, triphenyl phosphite. We kinetically isolate the reverse liquid-liquid transition from glass transition and crystallization with an extremely high heating rate of flash differential scanning calorimetry, and prove the reversibility and first-order nature of liquid-liquid transition. Our finding not only deepens our physical understanding of liquid-liquid transition but also will initiate a new phase of its research from both fundamental and applications viewpoints.
} 
\end{abstract}

\maketitle
%
%
\newpage
%

Even for a single-component substance, there can be more than two liquid states \cite{Debene,poole1997polymorphic,mishima1998relationship, tanaka2000general}. The transition between these different liquid states is called ``liquid-liquid transition (LLT)''.  LLT is one of the most mysterious phenomena in liquid science and its presence and absence have often been debated for various systems. Since this problem is of fundamental importance in our understanding of the liquid state, LLT has kept attracting considerable attention. 

The presence of LLT has been reported for both molecular systems (water \cite{mishima1998relationship,poole1992phase,loerting2006amorphous,mallamace2007evidence}, triphenyl phosphite (TPP) \cite{tanaka_review,tanaka2013importance,tanaka2004liquid,mosses2014order,krivchikov2016thermal}, n-butanol \cite{KuriButa}, and possibly D-mannitol \cite{zhu2015possible}) and atomic systems 
(sulphur \cite{brazhkin2003high,mcmillan2007polyamorphism}, phosphorus \cite{katayama2000first}, silicon \cite{Sastry01, Si}, germanium \cite{Bhat} and ${\rm Y_{2}O_{3}}$-${\rm Al_{2}O_{3}}$ \cite{McMillan}). 
Recently, LLT was also reported for metallic glass-formers \cite{wei2013liquid,stolpe2016structural}. 
However, none of these examples is free from controversy. 
The situation is more complicated for molecular liquids than for atomic liquids, since LLT always takes place in a supercooled state metastable against crystallization for molecular systems \cite{tanaka2000general}.  
For atomic systems, on the other hand, the situation is better, since LLT often takes place in an equilibrium liquid state: For example, it was shown that liquid P shows a first-order like transition from P4 tetrahedra to polymereric P chain structure \cite{katayama2000first} and liquid S transforms into different polymeric structures upon heating \cite{brazhkin2003high,mcmillan2007polyamorphism}.  

One of the hottest and long-standing debates is on the nature of an unconventional amorphous state called ``glacial phase'' discovered by 
Kivelson and his coworkers \cite{cohen1996low} for a molecular liquid, triphenyl phosphite (TPP). 
We note that TPP is one of the most well-studied molecular systems which are expected to have LLT. 
When TPP is kept at a low temperature near but still above the glass transition temperature $T_{\rm g}$ ($\sim 204$ K), a supercooled state of liquid 1 slowly transforms to an apparently amorphous state 
distinct from its ordinary glass state. 
Since this transformation occurs above $T_{\rm g}$ of liquid 1, $T_{\rm g}^1$, the final amorphous state must not be a glass state of liquid 1, glass 1. 
We showed that this phenomenon can naturally be explained by LLT from liquid 1 to a glass state of liquid 2, glass 2 \cite{tanaka2004liquid}. 
In this scenario, thus, the glacial phase is glass 2. 
In the following, for simplicity, we use the term ``LLT'' to express the transition between liquid 1 and liquid 2/glass 2 without distinguishing whether liquid 2 is in a liquid or glass state.

Besides the LLT scenario, many other explanations were also proposed for the nature of the glacial phase; 
for example, the glacial phase was interpreted as a mixture of glass 1 and nano-crystals \cite{hedoux1998raman,hedoux1999mesoscopic,hedoux2001low,hedoux2002conversion,hedoux2002description,hedoux2004contribution,hedoux2006micro,baran2014polymorphism}, a liquid crystal or plastic crystal \cite{johari1997calorimetric}, and an unconventional crystal called defect-ordered crystal \cite{Alba}. 
X-ray diffraction data of the glacial phase formed at a very low temperature show only broad amorphous peaks and no 
sharp Bragg peaks, indicating the absence of distinct translational order in the glacial phase \cite{cohen1996low,hedoux1999mesoscopic,derollez2004structural,mei2004local,murata_xray}. 
This feature cannot be explained by the plastic-crystal scenario. 
The glacial phase prepared at a very low temperature does not exhibit distinct birefringence \cite{tanaka2004liquid,shimizu2014evidence,kobayashi2015time}, 
which cannot be explained by the liquid-crystal scenario. 
On the other hand, the glacial phase formed at a rather high temperature exhibits not only weak birefringence \cite{cohen1996low, shimizu2014evidence,kobayashi2015time} 
but also small Bragg peaks \cite{hedoux2004contribution}. 
This can naturally be explained by the presence of nano-crystals in the glacial phase.  
Then the question is whether the glacial phase is primarily glass 2 or just a mixture of glass 1 and nano-crystals whose size decreases with a decrease in the annealing temperature, at which the glacial phase is formed. 
In the latter scenario, the absence of the Bragg peaks in the glacial phase formed at a very low temperature is ascribed to an extremely small size of nano-crystals.

Such debates may also originate from the counter-intuitive impression about LLT. According to classical liquid-state theory \cite{hansen1990theory}, 
the liquid state can be described by a single order parameter, density $\rho(\mathbf{r})$. Provided that a liquid is in a random disordered state, 
 it is hard to accept the presence of two liquids with different densities intuitively. 
However, once we accept that we need an additional scalar order parameter besides density to describe the state of a liquid, LLT is no longer counter-intuitive and can be accepted naturally. 
On the basis of this idea and along the spirit of the pioneering works by Str{\"a}ssler and Kittel \cite{strassler1965degeneracy} and Rapoport \cite{rapoport1967model}, we proposed 
a two-order-parameter model of liquid-liquid transition \cite{tanaka1999two,tanaka2000general,tanaka_review}. 
In this picture, for example, LLT in atomic systems like sulphur and phosphorous \cite{katayama2000first,brazhkin2003high,mcmillan2007polyamorphism} 
can be explained by the distinct change in the locally favoured structures stabilized by chemical (or, covalent) bonding. 
Similarly, locally favoured structures can also be formed by directional hydrogen bonding for molecular liquids.  
According to our two-order-parameter model \cite{tanaka2000general,tanaka_review,tanaka2013importance}, the order parameter governing LLT is the fraction of locally favoured structures, $S$, and then LLT is regarded as a gas-liquid-like transition of the order parameter $S$: Liquid 1 is a gas-like state with low $S$ whereas liquid 2 is a liquid-like state with high $S$. 
Since locally favoured structures are created and annihilated independently, their number density is not conserved and thus $S$ is a non-conserved scalar order parameter. 
Our recent X-ray scattering study \cite{murata_xray} revealed that upon LLT of TPP locally favoured structures whose size is a few nm are formed and its number density monotonically increases with time, 
and accordingly liquid 1 and liquid 2 can indeed be differentiated by the fraction of locally favoured structures $S$. 
Here we note that liquid 1 is a stable high-temperature liquid and liquid 2 is a low-temperature liquid, which usually exists in a glass state (glass 2) for TPP . 
This supports the two-order-parameter model 
of LLT \cite{tanaka2000general,tanaka_review,tanaka2013importance}. However, the controversy has still remained due to the lack of direct experimental evidence for the presence of two liquid states and the 
reversibility of LLT.  More importantly, no one have succeeded in avoiding nano-crystallization so far, which is due to an intrinsic difficulty associated with 
the fact that all the LLTs reported for molecular liquids such as TPP and water take place in a supercooled state below the melting point.   

To firmly establish the LLT scenario, it is desirable to show the reversibility of LLT without suffering from any crystallization. 
For a heating rate less than 1 K s$^{-1}$, however, the reverse LLT is hidden behind crystallization, even if it exists, since crystallization takes place immediately after 
the glass 2-to-liquid 2 transition during heating \cite{tanaka2004liquid}. 
Terashima et al. observed an endothermic (heat absorbing) peak upon heating, which they attributed to the reverse LLT on the basis of the heating rate dependence of the onset temperature of the peak \cite{terashima2013observation}. However, if this is the reverse LLT, we should also observe a glass 2-to-liquid 2 transition during heating before glass 2 goes back to liquid 1, since the process of the reverse LLT should occur only in a liquid state and not in a glass state. 
But they reported only one endothermic peak. 
Furthermore, with a slow heating rate employed in the previous studies, it is impossible to access the entire reverse process from the glacial phase to liquid 1 due to the interference by crystallization. 
Thus, it is still unclear whether the peak is due to the melting of nano-crystals, the glass 2-to-liquid 2 transition, and/or the reverse LLT process. 

In this Article, we aim at 
not only showing the reversibility of LLT but also confirming the coexistence of the two `liquid' states in the course of the transition in an unambiguous manner. 
These are crucial for proving that the transition takes place between two distinct liquid states. 
To this end, we apply ultra high-speed (flash) DSC 
(Differential Scanning Calorimetry), 
which can provide a heating rate more than 4 orders of magnitude higher than that of conventional DSC (see Methods). 
This allows us to avoid crystallization upon heating and to directly access the reverse process of LLT without the interference by crystallization. 
 
\section*{Results}
\subsection*{Overall transition behaviours}
\begin{figure}
\includegraphics[width=9cm]{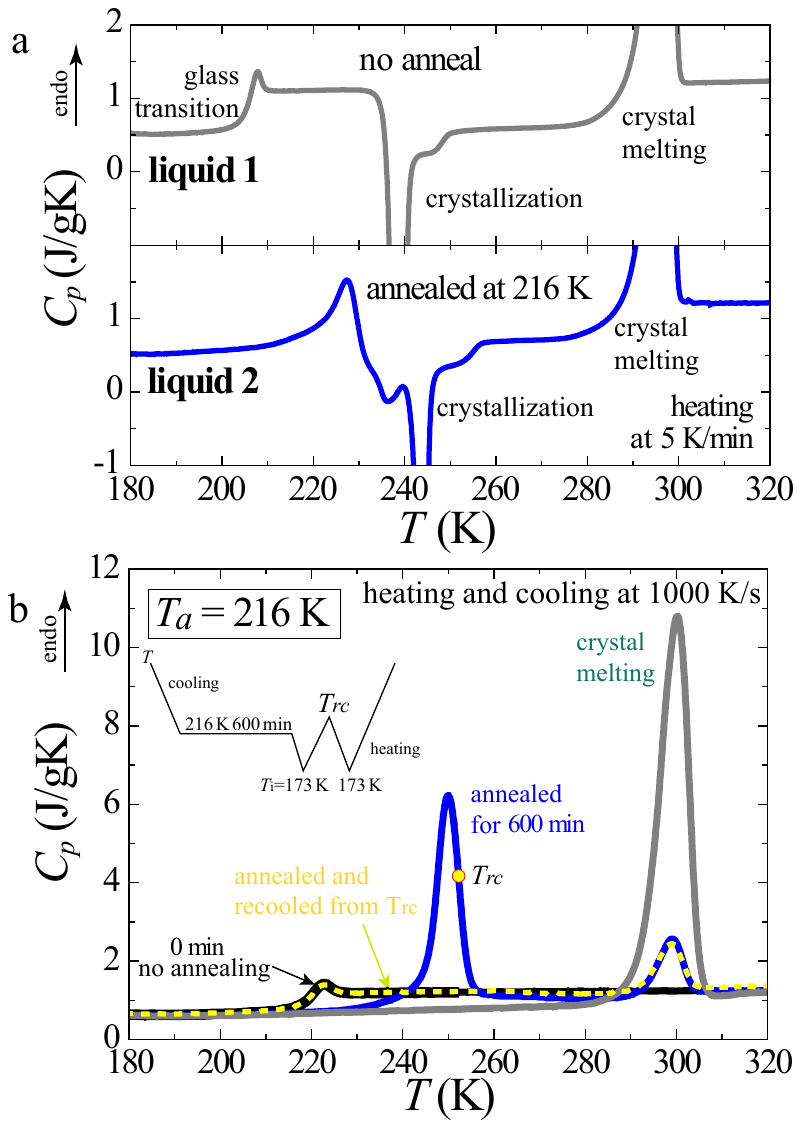}
\caption{
{\bf Comparison of DSC heat flow curves on heating between slow (1/12 K s$^{-1}$) and fast (10$^3$ K s$^{-1}$) rates.}
{\bf a,} The results for the slow heating rate. 
The grey curve is a heating curve of liquid 1 without annealing 
and the blue curve is a heating curve of liquid 2 obtained after annealing for 600 min at 216 K. 
After the complete transformation from liquid 1 to the glacial phase (glass 2) by constant-temperature annealing, 
the glass 1-to-liquid 1 transition signal completely disappears and instead there appears an endothermic peak at higher temperature, which is  
then followed by the significant exothermic peak due to crystallization. This exothermic peak makes it difficult to clarify the origin of the endothermic process. 
{\bf b,} The results of flash DSC measurements. The black curve is obtained for a sample without annealing (liquid 1) and 
the blue curve is for a sample after annealing (the glacial phase, or glass 2). 
The yellow dashed curve is taken after re-cooled from a point $T_{\rm rc}$ in the endothermic peak (see the inset for the temperature protocol).  
The glass transition signal of liquid 1 is observed in the yellow dashed curve around 220~K, indicating that the glacial phase (glass 2) 
has already returned to liquid 1 during the endothermic process before reaching $T_{\rm rc}$. The grey curve is for a sample fully crystallized. 
}
\label{fig:figure1}
\end{figure}

First we show calorimetric data obtained by the conventional DSC with a heating rate of 1/12~K s$^{-1}$ 
in Fig.~\ref{fig:figure1}a (see Methods). 
The grey curve in the top panel is a DSC curve for liquid 1, which is obtained without annealing 
after liquid 1 is vitrified into its glassy state, glass 1. There we can see the glass 1-to-liquid 1 transition, whose onset is located around 204~K. 
The signal also has a large exothermic (heat releasing) peak due to crystallization around $\sim 240$ K. 
On the other hand, the blue curve in the bottom panel shows a DSC heating curve for the glacial phase, 
which is prepared by annealing TPP for 600 min at 216 K until the transition is completed. 
We can see a change suggestive of the glass 2-to-liquid 2 transition, which starts around 210 K upon heating  
but is immediately followed by reverse LLT from liquid 2 to liquid 1 and crystallization (see below). 
The glass 2-to-liquid 2 transition is broader than the glass 1-to-liquid 1 transition, suggesting 
liquid 2 is less fragile than liquid 1 \cite{tanaka2004liquid}.    

Next we show DSC results obtained by the flash DSC with a heating rate of 10$^3$~K s$^{-1}$ in Fig. \ref{fig:figure1}b (see Methods). 
The black curve is a DSC curve for liquid 1, which was obtained without annealing 
immediately after liquid 1 is vitrified into its glassy state, glass 1, 
as in the top panel of Fig.~\ref{fig:figure1}a. 
$T_{\rm g}^1$ observed with a heating rate of  10$^3$~K s$^{-1}$ ($\sim 217$ K) is significantly higher than that observed with a slower heating rate of 1/12~K s$^{-1}$ 
($\sim 204$ K) by the conventional DSC (compare the black curve in Fig. \ref{fig:figure1}b with the grey curve in the top panel of Fig.~\ref{fig:figure1}a). 
This is consistent with the general rule that the glass transition temperature increases with an increase in the heating rate.  
As shown in Fig.~\ref{fig:figure1}b (see the black curve), 
the DSC signal obtained with the ultra-high speed heating exhibits no exothermic heat due to crystallization during the heating process, 
unlike the case of the slow heating in Fig.~\ref{fig:figure1}a (see the grey curve). Accordingly, there is no signature of crystal melting for a non-annealed sample (the black curve) in Fig. \ref{fig:figure1}b, 
which is supposed to occur around 300~K (see the grey curve in Fig.~\ref{fig:figure1}a). 
This clearly shows that the ultra-high speed heating successfully avoids the occurrence of crystallization after 
the glass 1-to-liquid 1 transition. 
After TPP is annealed at 216~K for 600~min, on the other hand, 
the glass transition signal of liquid 1 completely disappears and instead a large endothermic peak appears around 250~K (see the blue curve in Fig.~\ref{fig:figure1}b) upon heating. 
This indicates that 
there is no liquid 1 (or glass 1) left in the glacial phase. 
This fact cannot be explained by the nano-crystal scenario, since it assumes 
that the glacial phase is a mixture of glass 1 and nano-crystals 
(see Supplementary Note 1 and Supplementary Figures 1, 2, 3, 4, 5, and 6 
for further evidence against the nano-crystal scenario). 
Thus, we assign the glacial phase obtained by annealing to be glass 2.

\begin{figure}
\includegraphics[width=10cm]{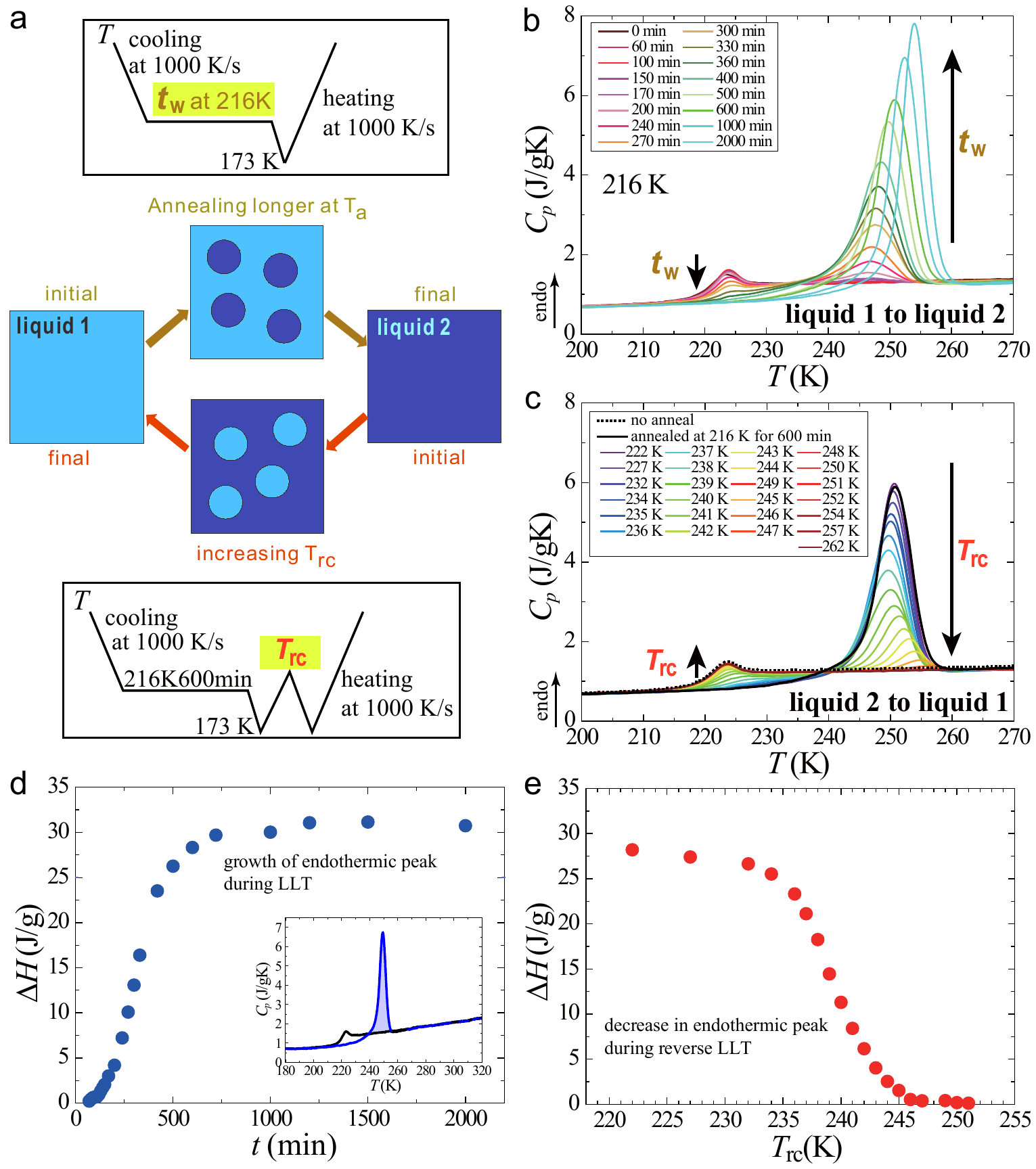}
\caption{{\bf Forward and reverse LLT processes.} 
{\bf a,} Schematic figures for forward LLT during annealing at $T_{\rm a}$ (top) and for reverse LLT as a function of the re-cooling  temperature $T_{\rm rc}$ (bottom). 
The experimental protocols for them are also show above and below the picture.  
{\bf b,} The forward LLT process from liquid 1 to liquid 2, probed by the reverse LLT process. 
The annealing time ($t_{\rm w}$-)dependence at $T_{\rm a}$=216~K. 
The numbers for lines with various colours denote $t_{\rm w}$. 
{\bf c,} The reverse LLT process from liquid 2 to liquid 1. 
The black dotted curve is for the sample without annealing 
and the black solid curve is for the sample annealed 
for 600 min at 216~K. 
The other curves are obtained for the second heating (see the protocol). 
The numbers for lines of various colours denote $T_{\rm rc}$. 
{\bf d,} The annealing time ($t_{\rm w}$-)dependence of the heat released by the reverse LLT upon heating for a sample annealed 
at 216 K. 
The inset illustrates how to estimate the transition heat by integration. 
{\bf e,} The re-cooling temperature ($T_{\rm rc}$-) dependence of the heat released upon heating for a sample annealed at 216 K for 600 min, in which 
LLT is completed. We can see that the onset of the reverse LLT, at which the endothermic heat starts to decrease,  is located around 235 K. 
This temperature is located slightly above $T_{\rm g}^2$. 
}
\label{fig:figure2}
\end{figure}

In order to clarify what is happening at the endothermic peak, 
we employ the following special temperature protocol  (see the inset in Fig.~\ref{fig:figure1}b and the bottom part of panel a of Fig. \ref{fig:figure2}). 
First we anneal a sample at 216~K for 600~min, which completely transforms liquid 1 to the glacial phase, and then quench it 
to $T_i=173$ K  
below $T_{\rm g}$. Next we heat the system until the temperature $T_{\rm rc}$ indicated by the yellow point on the blue curve, 
keep it at $T_{\rm rc}$ for a period of 0.1 s, and then cool it again from $T_{\rm rc}$ to $T_i$ 
= 173 K  
below $T_{\rm g}$.  
Here we use the cooling and heating rate of 10$^3$ K s$^{-1}$. 
The second heating from $T_i$ provides the yellow dashed DSC curve in Fig.~\ref{fig:figure1}b. 
We can clearly see the glass 1-to-liquid 1 transition signal in the yellow dashed curve. 
Furthermore, the perfect overlap of the glass transition signal between the black curve and the yellow dashed curve in Fig.~\ref{fig:figure1}b suggests that 
the glacial phase (or, glass 2) fully returns back to liquid 1 already much before crystal melting takes place in the heating process (more specifically, 
either before reaching $T_{\rm rc}$ in the first heating process or during 0.1 s kept at $T_{\rm rc}$). 
Thus, the endothermic peak around 250 K in the blue curve should not be associated with the crystals that should melt around 300 K. 

We can see that the crystal melting behaviour around 300~K is almost perfectly the same between the blue curve and the yellow dashed curve (see Fig.~\ref{fig:figure1}b).  
This result clearly indicates that the crystals are formed exclusively during annealing 
and they are not affected by heating and cooling below $T_{\rm rc}$, suggesting that liquid 2 is prone to crystallization 
compared to liquid 1.  
We also note that the amount of the heat of fusion in the blue curve is much smaller than that in the grey curve, which is for a fully crystallized sample.   
This can be explained as follows: 
crystallization takes place preferentially in liquid 2/glass 2 domains, which are newly formed during annealing, but its glassy nature inhibits both nucleation and growth of crystals. 

\subsection*{The LLT scenario}

We show experimental results on the forward and reverse LLT processes in much more detail (see the top part of 
Fig.~\ref{fig:figure2}a 
for the protocol and the resulting phase change process as a function of $t_{\rm w}$). 
The annealing time $t_{\rm w}$-dependence of the first heating curve is shown in Fig.~\ref{fig:figure2}b. 
The glass 1-to-liquid 1 glass transition signal becomes smaller with an increase 
in $t_{\rm w}$ and completely disappears for $t_{\rm w} \geq 400$ min, indicating that the liquid 1-to-glass 2 transition is completed around this annealing time. 
On the other hand, a new endothermic peak appears for $t_{\rm w} \geq 200$~min and continues to grow with an increase in $t_{\rm w}$. 
Another important fact is that after the endothermic peak the heat capacity of the liquid is the same as that of liquid 1, 
which can be seen from the fact that above 260 K all curves almost coincide with each other with the curve of $t_{\rm w}=0$ for liquid 1.  
This clearly indicates that the endothermic peak is associated with the transition from glass 2 to liquid 1. 
Furthermore, we can see in Fig.~\ref{fig:figure2}b that the endothermic peak position shifts to a higher temperature with an increase in $t_{\rm w}$. 
Figure \ref{fig:figure2}d shows the $t_{\rm w}$-dependence of the total heat released by reverse LLT, which should be proportional to the amount of liquid 2 formed during $t_{\rm w}$. 

We also show the $T_{\rm rc}$-dependence of the second heating curve in Fig.~\ref{fig:figure2}c (see the bottom part of Fig. \ref{fig:figure2}a 
for the protocol and the schematic figure showing the resulting phase change process as a function of $T_{\rm rc}$).   
Here the sample is kept for 0.1 s at $T_{\rm rc}$ before re-cooling from $T_{\rm rc}$.  
With an increase in $T_{\rm rc}$, the endothermic peak becomes smaller and the glass 1-to-liquid 1 transition signal emerges and gradually becomes larger. 
Figure \ref{fig:figure2}e shows the $T_{\rm rc}$-dependence of the heat released during the reverse LLT, which should be proportional to the amount of liquid 2 remaining after heated to $T_{\rm rc}$. 

On the basis of these results, we discuss the origin of the new endothermic peak emerging after annealing at 216 K 
(see the blue curve in Fig.~\ref{fig:figure1}b). 
There are the following three possible origins for the endothermic peak appearing around 250 K: (i) the glass 1-to-liquid 1 transition, (ii) the glass 2-to-liquid 2 transition, and 
(iii) the reverse LLT from liquid 2 to liquid 1. 
Whichever the transformed state contains liquid 1 or liquid 2, the system is initially in a glassy state and, thus, it should exhibit a glass transition signal upon heating before finally returning to 
liquid 1 \cite{cohen1996low}. 
There is a difference in the heat flow level between before and after the endothermic peak, indicating the difference in the heat capacity $C_p$ (see the blue curve in Fig.~\ref{fig:figure1}b). 
This is consistent with the occurrence of glass transition. 
However, we show below that the glass transition cannot be a primary origin of the endothermic peak. 

Firstly we consider possibility (i) that the endothermic peak is due to the glass 1-to-liquid 1 transition. 
Although the position of the endothermic peak is significantly different from the glass transition peak of liquid 1 ($t_{\rm w}$ = 0), it alone does not immediately mean 
that the system is liquid 2 and not liquid 1. 
This is because ageing can generally shift the glass transition peak towards a higher temperature. 
Thus, even if the peak is due to the glass 1-to-liquid 1 transition, the peak position can depend on $t_{\rm w}$: 
the ageing of glass 1 should continuously shift the peak towards a higher temperature and increase the magnitude of the glass transition signal.    
Contrary to this expectation, however, Fig.~\ref{fig:figure2}b tells us that an increase in $t_{\rm w}$ reduces the signal of the glass-to-liquid transition and leads to the emergence of a new endothermic peak 
at a much higher temperature and the increase of its hight. 
This observation indicates that there are clearly two transitions with different origins. 
Thus, the presence of the two distinct transitions cannot be explained 
by scenario (i) based on the ageing of glass 1. 
This conclusion is also supported by the fact that after the transition ($t_{\rm w}>400$ min) the glass transition signal associated with liquid 1 component completely disappears.   

Next we consider possibility (ii) that the endothermic peak is mainly due to the glass 2-to-liquid 2 transition. 
The step-like change at a lower temperature is definitely associated with the glass 1-to-liquid 1 transition at least for a rather short annealing time $t_{\rm w}$.  
On the other hand, the endothermic peak appearing after annealing should be associated with liquid 2 formed during annealing. 
We note that the temperature shift of the peak towards a high temperature with an increase in $t_{\rm w}$ does not stop even after the transition is completed. 
However, the total heat released during the transition becomes constant after the completion of the transition ($t_{\rm w}>1000$ min), as shown in Fig.~\ref{fig:figure2}d. 
The ageing of a glass should lead to the simultaneous increase in both the transition temperature and peak area. 
The lack of this feature indicates that the heat involved in the endothermic peak cannot be explained by the glass 2-to-liquid 2 transition alone, even taking the ageing effect into account. 
Furthermore, as shown in Fig. \ref{fig:figure1}b, glass 2 has already returned to liquid 1 at $T_{\rm rc}$ upon heating, the endothermic peak 
should involve the reverse LLT (see also below for the further supporting evidence). 
Thus we conclude that the endothermic peak is primarily not due to the glass 2-to-liquid 2 transition, although it should contribute partially.  

Finally, we consider the remaining possibility (iii) that the endothermic peak should come mainly 
from the reverse LLT from liquid 2 to liquid 1. 
This scenario is strongly supported not only by the above-mentioned transformation of glass 2 to liquid 1 before reaching $T_{\rm rc}$ but also 
by the fact that the heat released by the reverse LLT ($\sim 30$ J/g)  (see Fig. \ref{fig:figure2}d,e) 
is comparable to the heat absorbed by the forward LLT ($\sim$ 25-27 J/g).  
Here we note that the transition heat of the reverse LLT does not depend on 
the heating rate in the range of 500 $\sim$ 2000 K s$^{-1}$  
and is almost constant within $\pm$ 1 \%. 
This reflects that the transition is between the well-defined glass 2 state, which is almost uniquely determined by the 
annealing temperature and the annealing time, and the liquid 1 state. 
The difference in the transition heat between the reverse and forward LLT 
may come from the contribution of the glass 2-to-liquid 2 transition. 
The onset of the glass 2-to-liquid 2 transition marks the onset of the heat release (see below on the details of the glass transition behaviour).  
We note that the glass 2-to-liquid 2 transition provides the system with mobility, which is necessary for the reverse LLT to proceed. 

In this reverse LLT scenario, we can explain the shift of the peak position towards a higher temperature with an increase in $t_{\rm w}$ as a consequence of the 
ageing of glass 2: Glass 2 becomes more stable and its glass transition temperature becomes higher with $t_{\rm w}$, 
leading to the shift of the onset of the reverse LLT towards a higher temperature. 
During annealing at $T_{\rm a}$, a system gradually transforms from liquid 1 to the glass state of liquid 2 (glass 2) with $t_{\rm w}$. 
Reflecting this, the total heat released during the reverse LLT should increase with an increase of $t_{\rm w}$. For $t_{\rm w}>1000$ min, however, 
it becomes constant since LLT is completed, i.e., the system almost perfectly becomes glass 2 (see Fig. \ref{fig:figure2}b,d). 
This saturation indicates that the order parameter $S$ in glass 2 becomes almost constant for $t_{\rm w}>1000$ min. 
So we conclude that the transition behaviour shown in Fig.~\ref{fig:figure2}b consists of the step-like glass 2-to-liquid 2 transition   
and the endothermic peak due to the reverse LLT from liquid 2 to liquid 1 (see below on the separation of the two transitions). 

Next we consider the experimental results using the special temperature protocol (see the bottom part of Fig. \ref{fig:figure2}a for the protocol and Fig. \ref{fig:figure2}c for the results). 
In the above, we show that liquid 2 already fully returns to liquid 1 at the yellow point marked on the blue curve in Fig.~\ref{fig:figure1}b. 
Particularly, Fig.~\ref{fig:figure2}c shows that the glass transition signal associated with liquid 1 gradually recovers 
during the endothermic process with increasing $T_{\rm rc}$, clearly supporting the above-mentioned scenario that 
this endothermic peak is due to the reverse LLT process (liquid 2 $\rightarrow$ liquid 1). 
This is also consistent with the fact that the reverse LLT process from liquid 2 to liquid 1 should be an endothermic process 
since the forward LLT from liquid 1 to liquid 2 during isothermal annealing is an exothermic process 
\cite{cohen1996low,tanaka2004liquid}. 
We indeed confirm the amount of heat associated with the transition is about the same between the forward and reverse processes, as mentioned above. 
We can see in Fig.~\ref{fig:figure2}c that the peak position slightly shifts towards a higher temperature for higher $T_{\rm rc}$. 
This suggests that more stable parts of liquid 2 with higher $S$ transform to liquid 1 at higher $T_{\rm rc}$. 

\subsection*{Glass transition behaviour and its link to the type of LLT}
Now we focus on the glass transition behaviour taking place prior to the reverse LLT upon heating (see also Supplementary Note 2 and Supplementary Figures 7 and 8). 
The glass transition behaviours shown in Fig.~\ref{fig:figure2}b and c suggest that, in both processes of the forward and reverse 
LLT, the two disordered phases 
of liquid 1 and 2 coexist and transform reversibly with each other (see the schematic pictures in Fig. \ref{fig:figure2}a). 
This is the first unambiguous evidence not only for the reversibility and the first-order nature of LLT in molecular liquids but also for the fact that the transition is between two distinct liquid states. 
The presence of the two glass transitions and the direct reversibility of the transition between the two liquid phases can be naturally explained by the LLT scenario, but not by 
the nano-crystal scenario. 

\begin{figure}
\includegraphics[width=14cm]{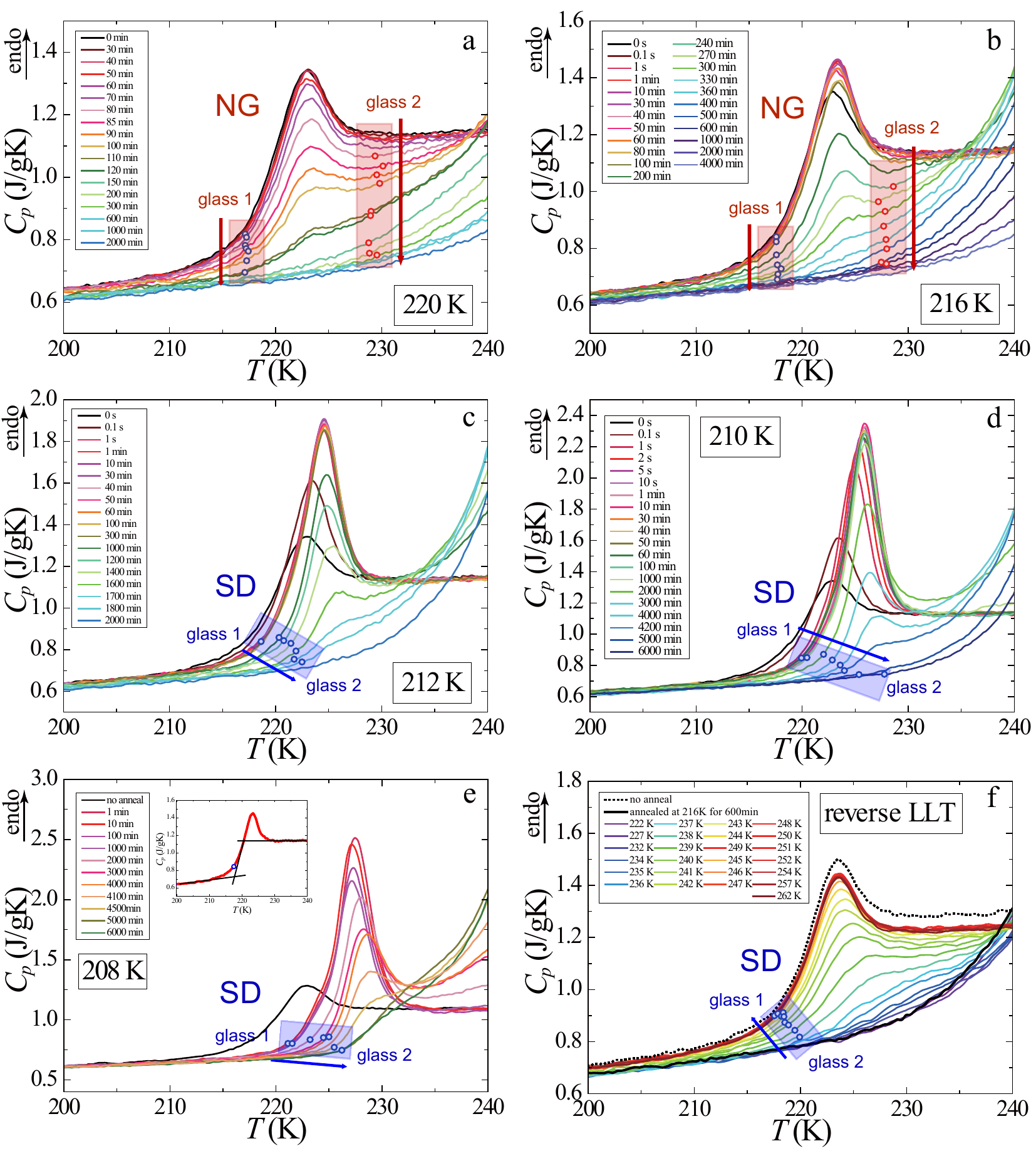}
\caption{{\bf Glass transition behaviours during the process of forward and reverse LLT.}
{\bf a-b,} The temporal change in the glass transition behaviour during NG-type LLT observed at $T_{\rm a}=220$ K and 216 K, respectively. 
{\bf c-e,} The same observed during SD-type LLT observed at $T_{\rm a}=212$, 210, and 208 K, respectively. 
{\bf f,} The dependence of the glass transition behaviour as a function of $T_{\rm rc}$ in the reverse LLT process. 
The open circles show the onset temperatures of glass transition and the widths of the half transparent belts roughly represent possible errors in their determinations. 
An example of the estimation of the onset temperature of a glass transition is shown in the inset of panel e.  
The arrows indicate the directions of the increase in $t_{\rm w}$ for panels a-e 
and that in $T_{\rm rc}$ for panel f.  
}
\label{fig:figure3}
\end{figure}

To be more quantitative, we identify the onset of the glass-to-liquid transition of the glass state obtained by various annealing time $t_{\rm w}$,  
as shown in Fig. \ref{fig:figure3} (see the inset of panel e on the determination of the onset of the glass transition). 
The glass transition behaviour provides crucial information on the type of LLT, i.e., whether LLT is nucleation-growth (NG)-type or spinodal decomposition (SD)-type \cite{tanaka2000general}. 
We note that NG-type LLT proceeds in a metastable state while accompanying nucleation of liquid 2 droplets in liquid 1, whereas SD-type LLT proceeds in an unstable state by a continuous transformation of liquid 1 to liquid 2. 

When LLT proceeds above $T_{\rm SD}^{1 \rightarrow 2} \sim 214$ K, we have two sequential 
glass transitions upon heating. For $T_{\rm a}$=216 K, for example, we identify $T_{\rm g}^1 \sim 217$ K and $T_{\rm g}^2 \sim 228$ K (see Fig. \ref{fig:figure3}b). 
Interestingly, the positions of the onsets of the two glass transitions do not depend upon the annealing time within errors ($\pm 2$ K). This is a characteristic feature of NG-type of LLT, 
reflecting that, for NG-type LLT, liquid 2 with the final order parameter value is nucleated in liquid 1 with the initial order parameter value 
and thus the order parameter changes discontinuously from that of liquid 1 to that of liquid 2. 
This coexistence of the two `liquid' phases during transformation can be regarded as a direct manifestation of LLT and its first-order nature \cite{tanaka2000general}.  
Here it is worth mentioning that the glass 2-to-liquid 2 transition behaviour is not so clear compared to the glass 1-to-liquid 1 transition. 
The spinodal temperature, $T_{\rm SD}^{2 \rightarrow 1}$,  
or the stability limit of liquid 2 against liquid 1 upon heating, is estimated to be around 235 K, 
which is located only slightly above the onset of the glass 2-to-liquid 2 transition. Thus, the glass transition does not complete at this temperature within a short time. 
This means that the plateau of $C_p$ after the glass 2-to-liquid 2 transition never appears, making it difficult to observe a typical glass transition signal. 
For the reverse LLT to take place, the system needs to gain mobility. Once the system starts to gain mobility due to the glass 2-to-liquid 2 transition, the reverse LLT is immediately initiated. 
In other words, the glass transition and the reverse LLT almost simultaneously take place, making clear separation between the glass 2-to-liquid 2 transition and the reverse LLT intrinsically difficult.  

When LLT proceeds below $T_{\rm SD}^{1 \rightarrow 2}$ ($\sim 214$ K), on the other hand, we can see that there is only one glass transition, whose onset temperature continuously and gradually shifts from 
that of glass 1 to that of glass 2, as shown in Fig. 
\ref{fig:figure3}c-e. We note that we analyse the data only after the ageing is completed (see Supplementary Figure 9 and Supplementary Note 2).  
This glass transition behaviour is a characteristic feature of SD-type LLT, where the order parameter changes continuously with time \cite{tanaka2000general}. 
Unlike the NG-type LLT, the reverse LLT starts immediately after the first glass transition step of the glass state, whose transition temperature is located between those of glass 1 and 2 
in the process of LLT. 
Here it may be worth explaining why we may conclude that there is only one glass transition:  Below $T_{\rm SD}^{1 \rightarrow 2}$ ($\sim 214$ K), 
the $C_p$ exhibits a minimum following the glass-transition peak, but its value 
is larger than the heat capacity of liquid 1  (see the curve for $t_{\rm w}=2000$ min at $T_{\rm a}$=210 K in Fig. \ref{fig:figure3}d and the curves for $t_{\rm w} \geq 2000$ min at 208 K in Fig. \ref{fig:figure3}e). 
Considering that liquid 2 is in a more ordered state than liquid 1, the heat capacity of liquid 2 is expected to be smaller than that of liquid 1. Thus, the $C_p$ minimum larger than that of liquid 1 should stem from  an additional contribution to $C_p$ from the reverse LLT. This means that the reverse LLT already starts before the completion of the glass 2-to-liquid 2 transition   
and there is no other glass transition. 
We note that such behaviour is never observed for the case of NG-type LLT (see Fig. \ref{fig:figure3}a and b). 
This simultaneous occurrence of the glass transition and the reverse LLT is further supported by a clear two-step feature in the DSC curves for $t_{\rm w} \geq 4500$ min at $T_{\rm a}$=208 K in Fig.~\ref{fig:figure3}e.  
The first step is the glass-to-liquid transition and the second one is the reverse LLT, unambiguously indicating that there are two distinct sequential transitions. 
Finally, we stress that these glass transition behaviours cannot be explained by the nano-crystal scenario, in which there should be only one glass transition 
from glass 1 to liquid 1 at $T_{\rm g}^1$. 

\begin{figure}
\includegraphics[width=10cm]{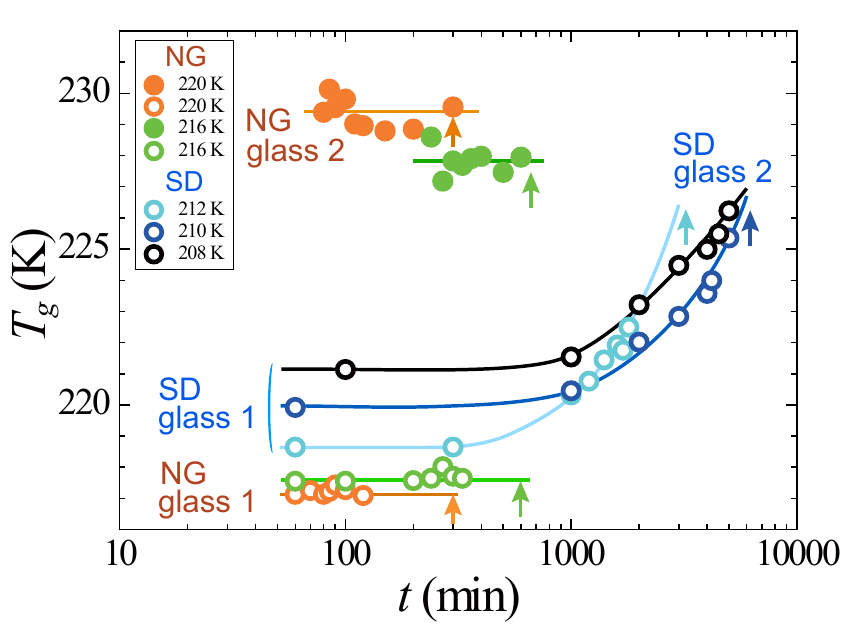}
\caption{{\bf The behaviour of the onset temperature of glass transition in the process of forward LLT.}
Each arrow indicates the time when the transition is completed for each annealing temperature. 
For NG-type LLT the onset temperatures of the two glass transitions (glass 1 and 2) are both constant with $t_{\rm w}$, whereas 
for SD-type LLT there is only one glass transition and its onset temperature continuously increases with $t_{\rm w}$. 
}
\label{fig:figure4}
\end{figure}

We compile all the data of the onset of the glass transition temperature in Fig. \ref{fig:figure4}. 
There we can clearly see that for SD-type LLT below $T_{\rm SD}^{1 \rightarrow 2}\sim 214$ K the onset temperature of the glass transition gradually 
and continuously shifts towards a high temperature with $t_{\rm w}$ whereas for NG-type LLT above $T_{\rm SD}^{1 \rightarrow 2}$ 
those of glass 1 and 2 stay almost constant as a function of $t_{\rm w}$ until the completion of the transformation, whose timing is indicated by the arrows in Fig. \ref{fig:figure4}. 
This is fully consistent with the phenomenology of NG- and SD-type phase transformation \cite{tanaka_review}.  

We stress that the above glass-transition behaviours are fully consistent with the characteristics of pattern evolution during LLT revealed by optical microscopy 
observation \cite{tanaka2004liquid}. 
The microscopy observation suffered from a criticism stemming from a resolution problem: People may suspect that continuous evolution of smooth density fluctuations 
observed for SD-type LLT may be merely a consequence of that droplets are actually formed but too small to be optically resolved. 
Our DSC results clearly indicate that this is not the case and SD-type LLT of the continuous nature indeed takes place below $T_{\rm SD}^{1 \rightarrow 2}$. 
Thus, our finding strongly supports the physical picture of the two-order-parameter model of LLT \cite{tanaka2000general}.

Finally, in the reverse LLT process, we can see that the onset temperature of the glass transition continuously shifts from that of liquid 2 to that of liquid 1, as shown in Fig. \ref{fig:figure3}f. 
This indicates that the reverse LLT takes place via SD-type transformation. This is consistent with the fact that the process takes place almost immediately (less than 0.1 s) 
without an incubation time. This fast transformation process implies that the transformation takes place in a liquid state with fast dynamics, i.e.,  
far above the glass transition from glass 2 to liquid 2, $T_{\rm g}^2$. 
The onset temperature below which the transition heat starts to decrease is located around $T_{\rm rc}\sim 235$ K (see Fig. \ref{fig:figure2}e). 
Thus, this $T_{\rm rc}$ marks the stability limit of liquid 2 against liquid 1 upon heating, i.e., the temperature above which liquid 2 becomes unstable against liquid 1, i.e., $T_{\rm SD}^{2 \rightarrow 1} \sim 235$ K. 
At ambient pressure, there is a rather large difference between $T_{\rm SD}^{1 \rightarrow 2}$ and $T_{\rm SD}^{2 \rightarrow 1}$, but this difference 
is expected to decrease with an increase in pressure and should disappear at the critical pressure $P_{\rm c}$. There 
the two spinodal temperatures $T_{\rm SD}^{1 \rightarrow 2}$ and $T_{\rm SD}^{2 \rightarrow 1}$ should merge to the critical temperature $T_{\rm c}$, which should be 
located above 235 K.   
From this, we can conclude that $T_{\rm SD}^{2 \rightarrow 1}$ is located around 235 K for $T_{\rm a}=216$ K. 
This can also be confirmed in Fig. \ref{fig:figure3}e (see the curve at $t_{\rm w}=4500$ min), where we can see the glass transition and reverse LLT separately. 
We note that it is located slightly above $T_{\rm g}^2$ ($\sim 228$ K for $T_{\rm a}=216$ K at a heating rate of $10^3$ K s$^{-1}$).

\section*{Discussion}
In summary, we reveal by ultra high-speed calorimetry that, upon heating, glass 2 that is formed by annealing liquid 1 
at a low temperature for a long time, first transforms into liquid 2 via the glass transition and then almost simultaneously liquid 2 becomes liquid 1 via the reverse LLT. 
Our study not only shows the reversibility of LLT of TPP, but also its first-order nature from 
the coexistence of the two distinct liquid phases during LLT and the transformability between them. 
Furthermore, the glass transition behaviours of an intermediate state formed during LLT upon fast heating tell us that there are two types of LLT, 
NG-type and SD-type LLT, which are typical non-equilibrium dynamical processes of the first-order phase transition respectively in its metastable and unstable state.  
We successfully reveal the discontinuous and continuous nature of the order parameter evolution for NG-type and SD-type LLT respectively 
from the glass transition behaviour of a transient state during LLT.  
This firm experimental confirmation of the first-order liquid-liquid transition in a single-component molecular liquid may initiate a new phase 
of theoretical and experimental research on the physical nature of this intriguing phase transition phenomenon and lead to a deeper understanding of the liquid state of matter. 
It may also contribute to the resolution of the controversy on LLTs of various systems such as water, which are also supposed to occur in a non-equilibrium metastable state  
\cite{loerting2006amorphous}: our experimental method may be useful for isolating LLT from nano-crystallization for other systems. 

\vspace{1cm}
\noindent
{\bf Methods.} 

{\bf Material. }
Triphenyl phosphite (TPP,  99.7\% purity) was purchased from Across organics and used it without further purification. 
The melting point $T_{\rm m}$ = 297~K, whereas the glass transition temperature of liquid 1 $T_{\rm g}^1$ = 204~K at a heating rate of 5 K min$^{-1}$. 

{\bf Calorimetry measurements.}
We used an ultra high-speed DSC (Mettler-Toledo Flash DSC 1) and a conventional DSC (Mettler-Toledo DSC 1). 
In ultra high-speed DSC measurements, 
the sample mass was 20-50 ng, which was estimated for each sample 
by comparing the heat of fusion of a fully crystallized sample obtained by the flash DSC with that obtained by the conventional DSC. 
The fully crystallized sample was obtained by 
heating at 20 K min$^{-1}$ from 173 K. 
In the forward and reverse LLT experiments,
we used the protocol shown in Fig.~\ref{fig:figure2}a, where the cooling and heating rate were 1000 K s$^{-1}$,  
the lowest temperature was 173 K, a waiting time of 0.1 s was inserted between each scan for stabilization of the instrument.
In conventional DSC measurements, 
the sample mass was 11.63 mg, the cooling and heating rates were 
10 K min$^{-1}$ and 5 K min$^{-1}$ respectively, and the lowest temperature was 173 K. 
All calorimetric measurements were performed under the N$_2$ atmosphere. 

{\bf Data availability.}
The data that support the findings of this study are available from 
the authors upon request.

%
%

\vspace{1cm}
\noindent
{\bf Acknowledgements}
\noindent
The authors thank C.A. Angell for valuable discussions. 
This study was partly supported by Specially Promoted Research (25000002), Scientific Research (S) (21224011), and Scientific Research (C) (16K05510) 
from the Japan Society for the Promotion of Science (JSPS). 

\clearpage
\setcounter{figure}{0}

\renewcommand{\figurename}{{\bf Supplementary Figure}} 
\renewcommand{\thefigure}{\arabic{figure}} 


\centerline{\bf \large Supplementary Information}
\begin{figure}[h!]
\includegraphics[width=12cm]{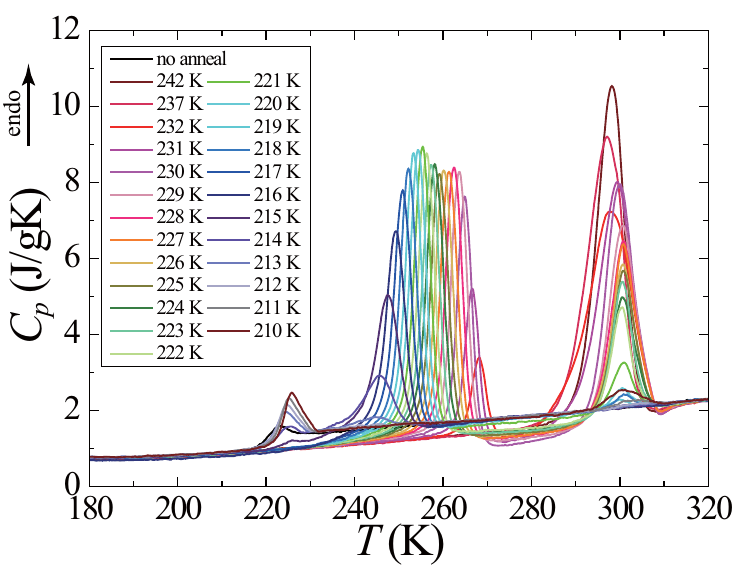}
\caption{
{\bf The DSC curve of TPP annealed isothermally for 600 min at various $T_{\rm a}$'s upon heating at 1000 K/s.}
Endothermic peaks around 240-270 K and melting peaks of bulk crystals around 300 K are separated, 
suggesting that the origin of endothermic peak is not due to the melting of nano-crystals (see text). 
}
\label{fig:tw600min}
\end{figure}
\clearpage

\begin{figure}[t!]
\includegraphics[width=8cm]{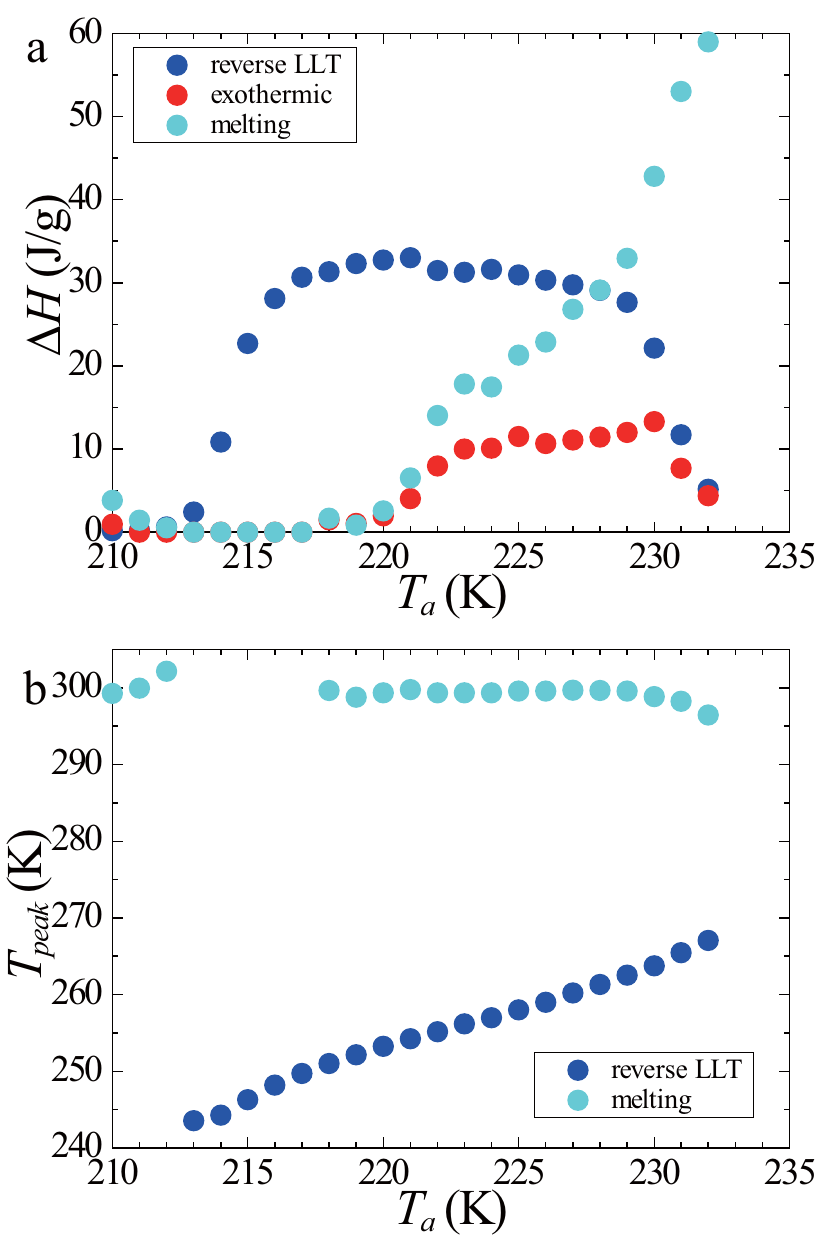}
\caption{{\bf Characteristics of the transitions of TPP  annealed isothermally for 600 min at various $T_{\rm a}$'s, upon heating at 1000 K/s.} 
{\bf a,} $T_a$-dependences of the transition heat of the reverse LLT, the exothermic signal appearing above $\sim270$ K due to crystallization, 
and the melting of crystal. 
{\bf b,} $T_a$-dependences of the peak temperatures for the reverse LLT and melting. 
}
\label{fig:tw600result}
\end{figure}
\clearpage

\begin{figure}[!h]
\includegraphics[width=9cm]{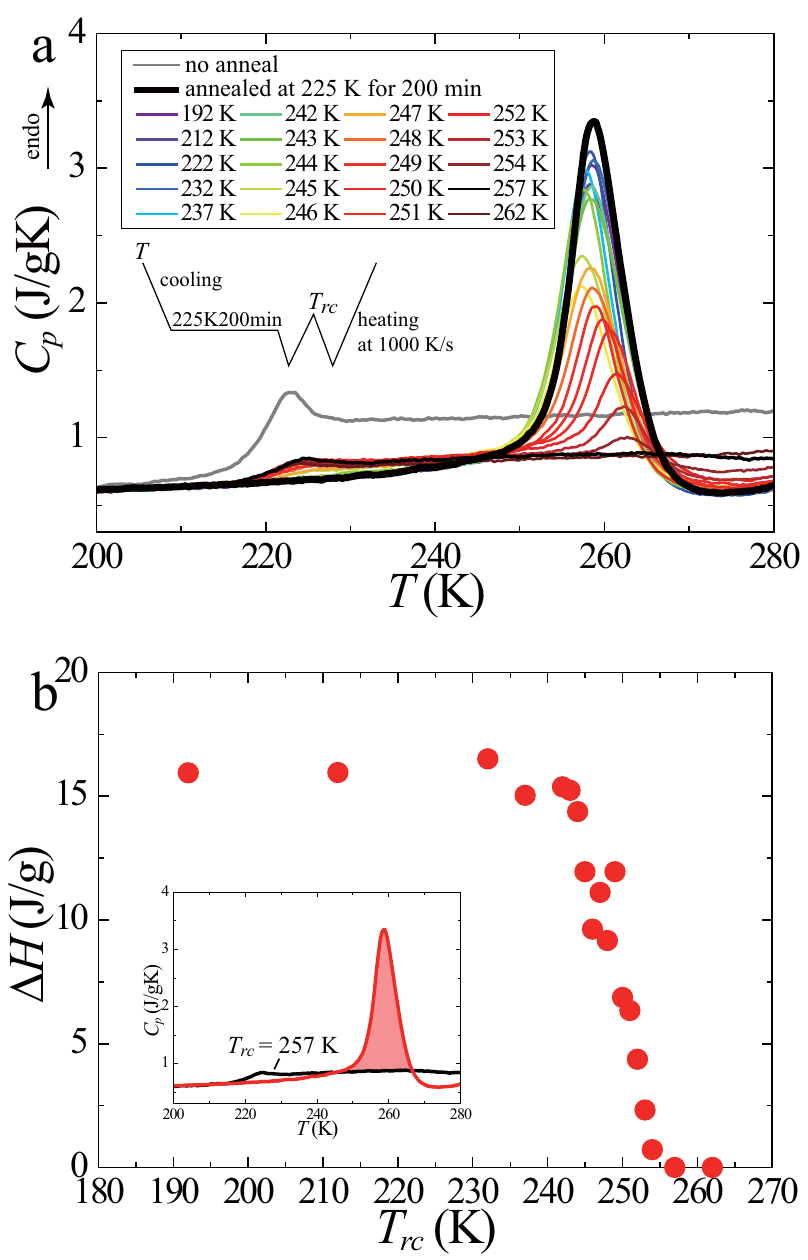}
\caption{{\bf The onset temperature of the reverse LLT for glass 2 formed at $T_{\rm a}=225$ K.}
{\bf a,} Reverse LLT processes upon second heating of samples, which are formed by annealing TPP at $T_{\rm a}$=225 K 
for 200 min, then heated to $T_{\rm rc}$, kept for 0.1 s there, and rapidly cooled to a low temperature. 
The temperature protocol and $T_{\rm rc}$ are shown in the inset. 
Note that only a part of the system returns to liquid 1 due to the crystallization. 
{\bf b,} $T_{rc}$-dependence of the heat released upon heating estimated 
from the results in panel a. 
The inset explains how to estimate the total heat released. 
The onset of the reverse LLT is found to be located around 242 K. 
}
\label{fig:Trc}
\end{figure}
\clearpage

\begin{figure}[t]
\includegraphics[width=10cm]{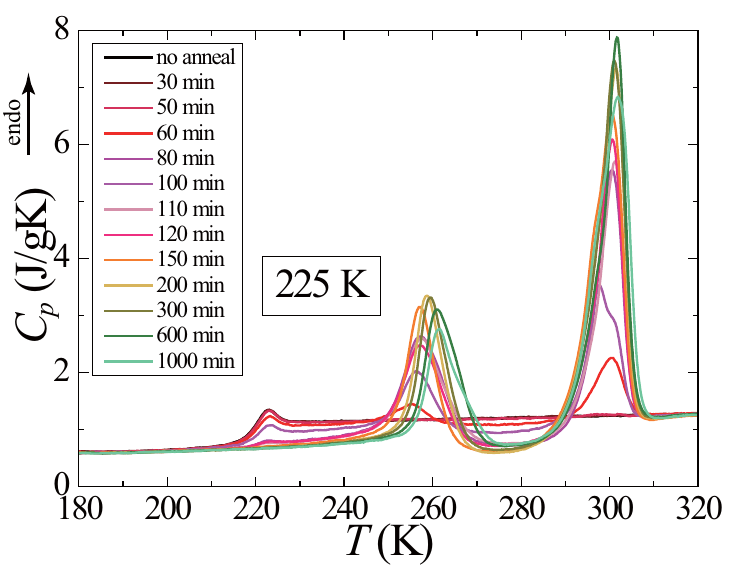}
\caption{
{\bf Annealing-time dependence of DSC heating curves of TPP annealed at 225~K.} 
The heating rate used in these experiments was 1000 K/s. 
}
\label{fig:225K}
\end{figure}
\clearpage

\begin{figure}[h!]
\includegraphics[width=15cm]{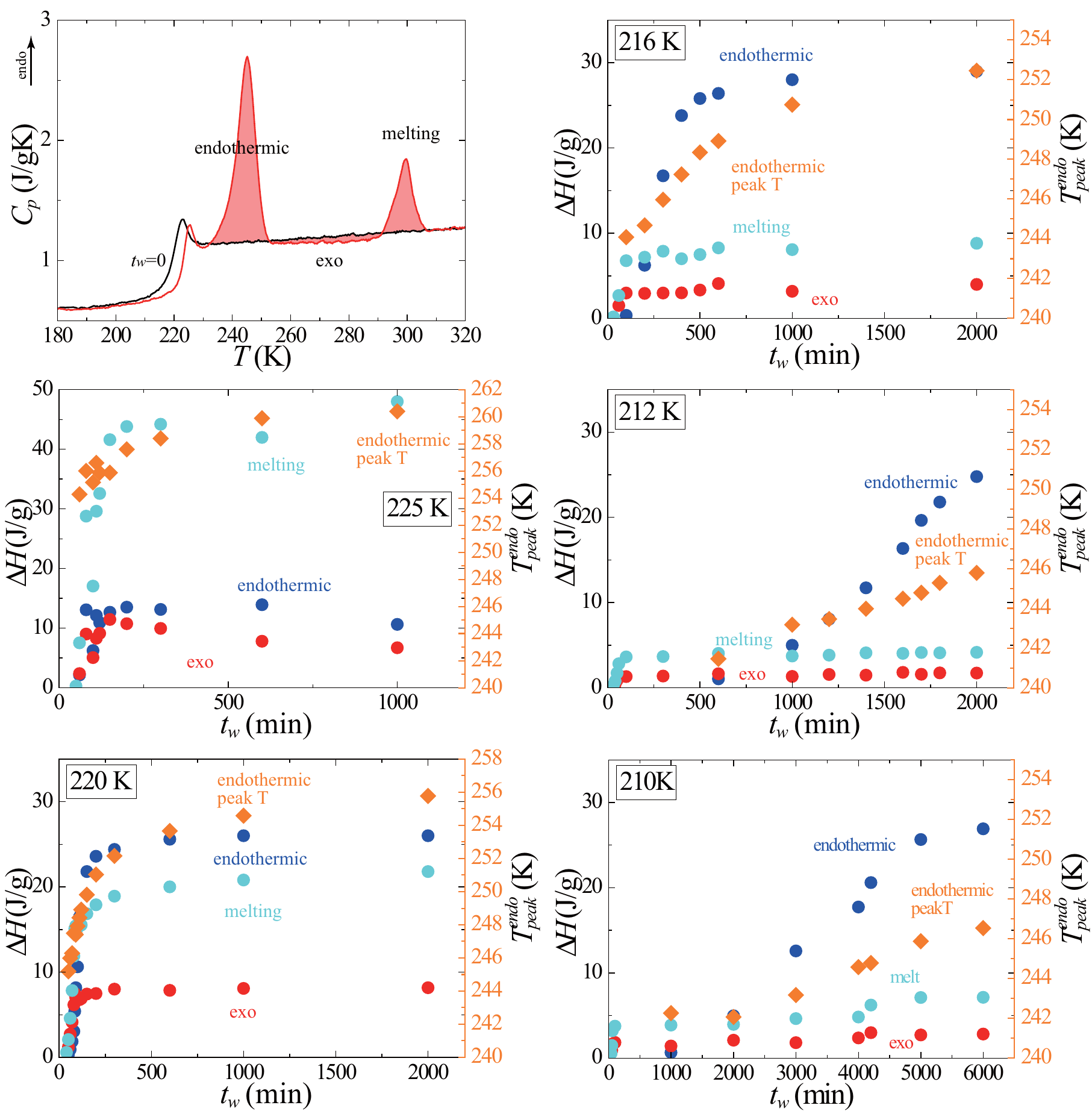}
\caption{
{\bf The time evolution of the transition heat $\Delta H$ during LLT for various $T_{\rm a}$'s.} 
We calculate $\Delta H$ by integrating the area of each heat process. 
We use a signal level at $t_w$ = 0 as the base line of the integration and 
an example of our analysis is shown in the first top left panel. 
}
\label{fig:deltaE}
\end{figure}
\clearpage

\begin{figure}[h!]
\includegraphics[width=9cm]{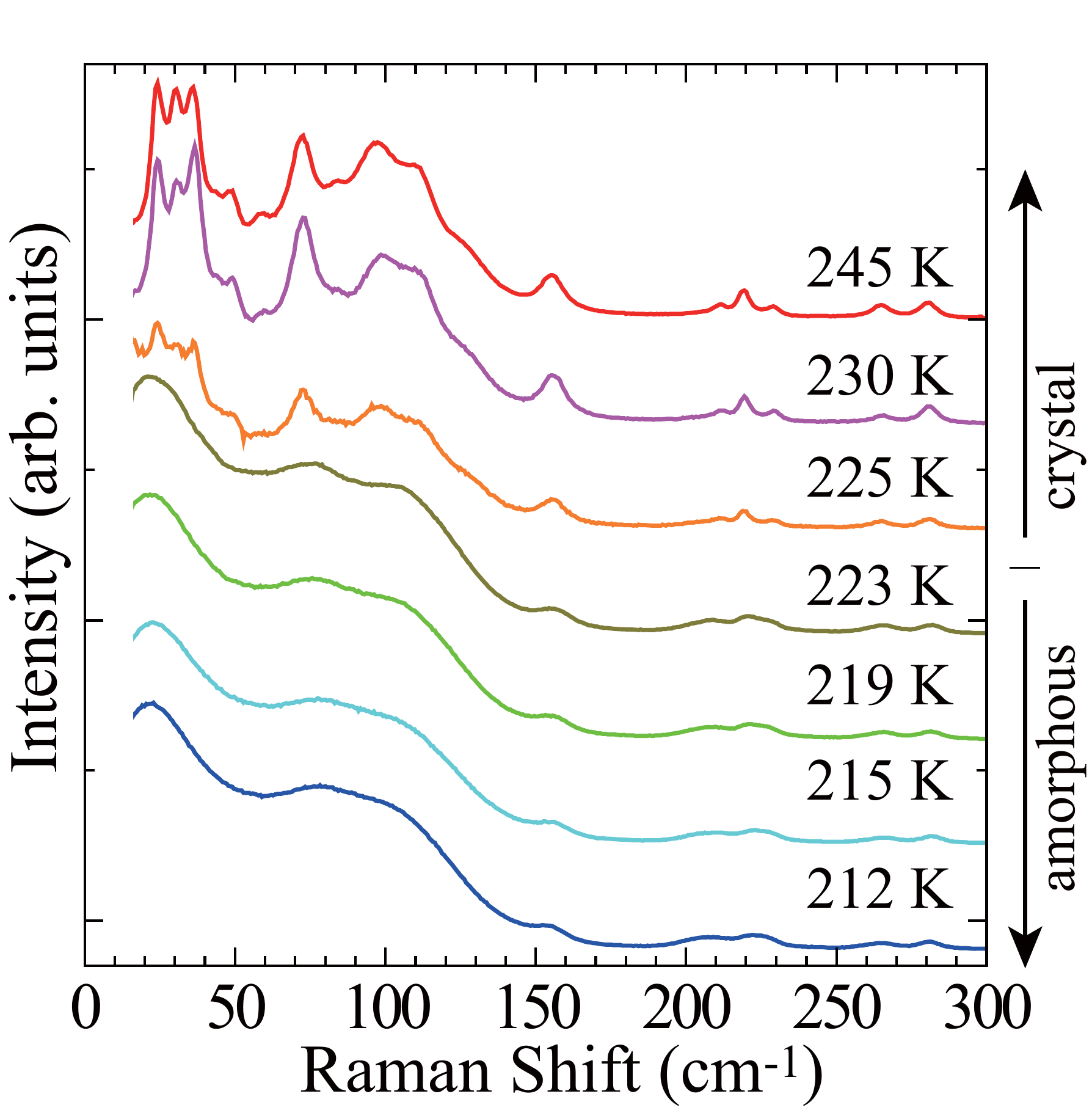}
\caption{
{\bf The annealing temperature $T_a$ dependence of Raman spectra.} We use the polarized incident laser light (532 nm) for excitation 
and detect the scattered light of all polarizations. All measurements are made after LLT is completed. 
The signals above 225 K indicates the presence of crystals, whereas those below 223 K are typical amorphous Raman spectra 
and do not show any indication of the presence of crystals. 
}
\label{fig:raman}
\end{figure}
\clearpage

\begin{figure}[b!]
\includegraphics[width=13cm]{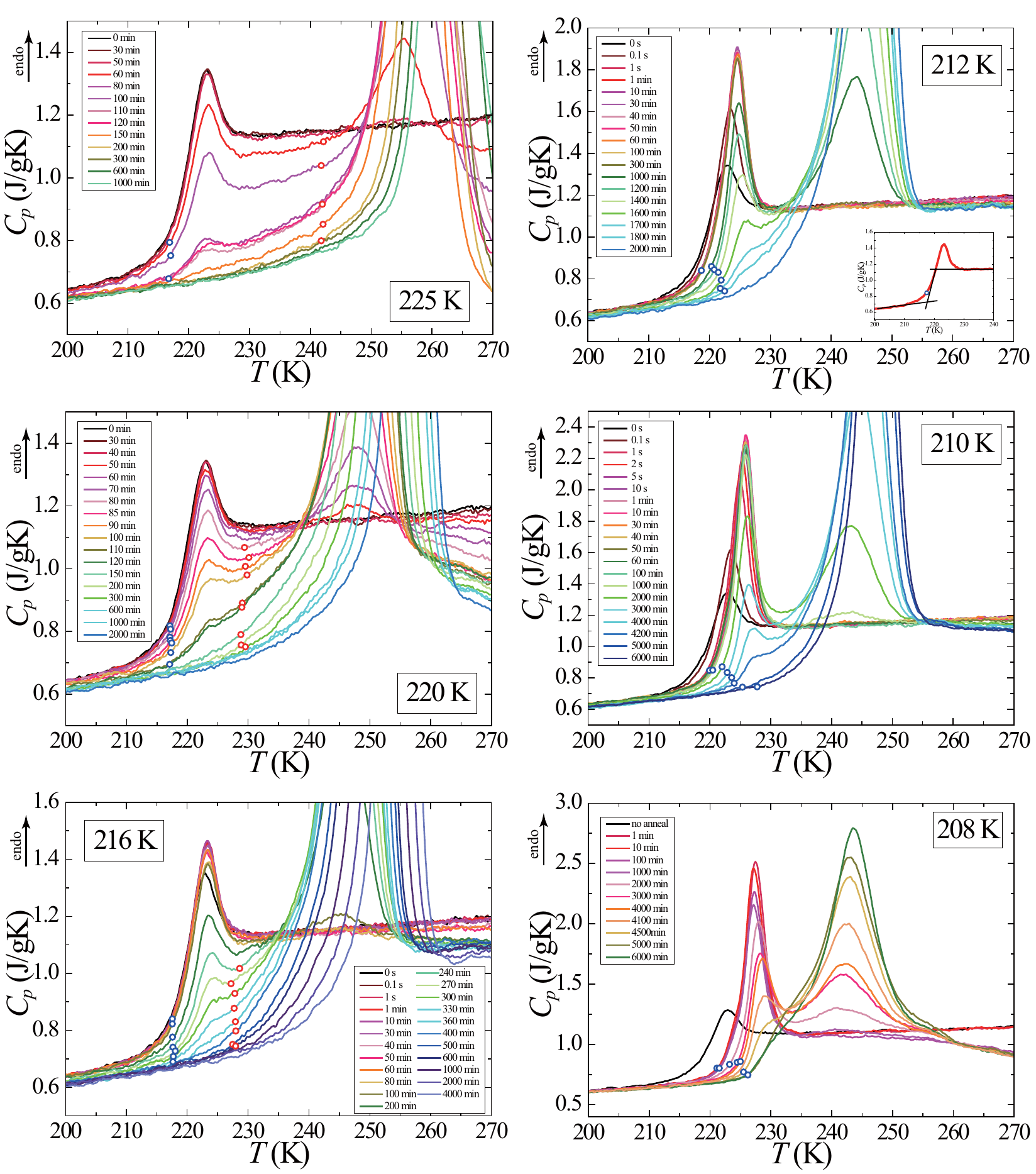}
\caption{
{\bf Annealing-time dependence of the glass transition and reverse LLT behaviours for six annealing temperatures.} 
Open circles denote the onset temperatures $T_{g}$. 
The $T_{\rm g}^1$ (blue circles) and $T_{\rm g}^2$ (red circles) do not change as a function of $t_{\rm w}$ above 214 K (NG-type), whereas 
$T_{\rm g}$ (blue circles) continuously changes below 214 K (SD-type). 
The inset in the panel of 212 K shows how to determine $T_{g}$. 
}
\label{fig:onsetTg1-all}
\end{figure}
\clearpage

\begin{figure}[b!]
\includegraphics[width=9cm]{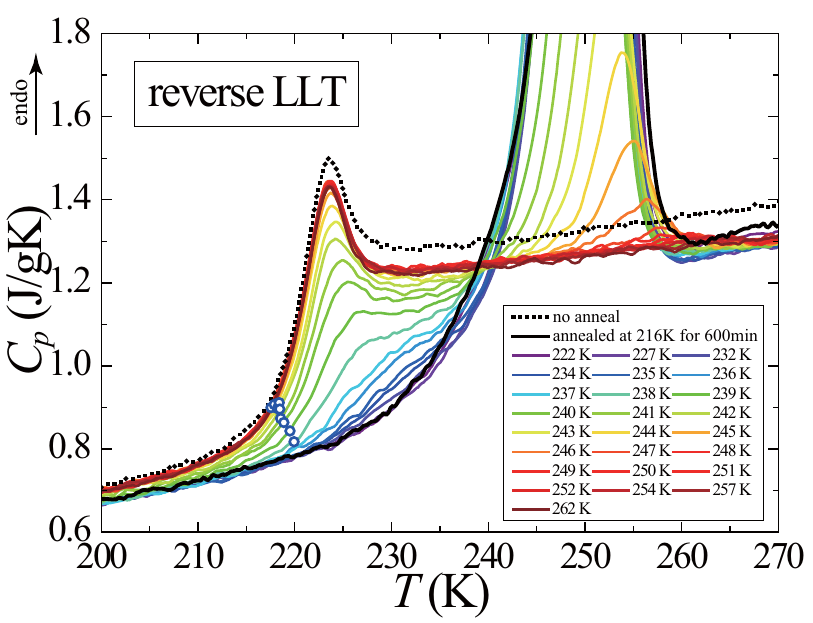}
\caption{
{\bf  $T_{\rm rc}$-dependence of the glass transition behaviours during the reverse LLT.} 
In this case, we also see the gradual continuous change of $T_{\rm g}$ during the reverse LLT transition, 
which is suggestive of the SD-type nature of the transformation. 
}
\label{fig:reverse}
\end{figure}
\clearpage

\begin{figure}[b!]
\includegraphics[width=9cm]{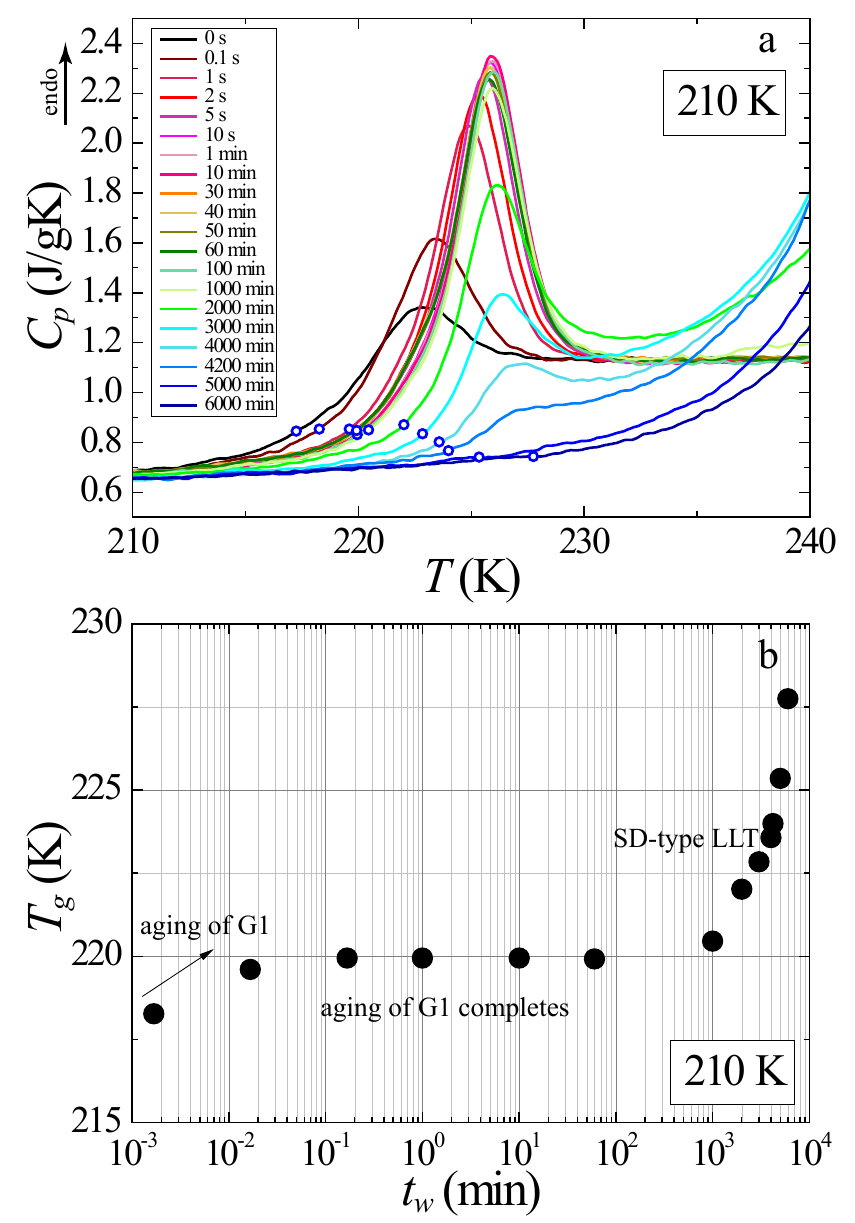}
\caption{
{\bf Temporal change in the glass transition behaviour during SD-type LLT at 210 K.} 
{\bf a,} The temporal change in the glass transition behaviour observed at 210 K and the estimated onset temperatures 
of the glass transition (open circles). 
{\bf b,} The time evolution of the onset temperature of the glass transition for a sample annealed at 210 K. 
}
\label{fig:ageing}
\end{figure}
\clearpage

\noindent
{\bf Supplementary Note 1: 
Difficulties of the nano-crystal scenario of the glacial phase}

There have been long-standing arguments that the transition we focus here is nano-crystal formation \cite{hedoux1998raman,hedoux1999mesoscopic,hedoux2001low,hedoux2002conversion,hedoux2002description,hedoux2006micro,baran2014polymorphism}. In this scenario, the endothermic peak 
observed upon heating of the glacial phase should be due to the melting of nano-crystals. Although the endothermic peak is located at a temperature 
much lower than the melting point of bulk crystal, this might be explained by very small sizes of nano-crystals 
and their defective structures \cite{hedoux2002conversion,Alba}. 
If we assume this, the system can have two melting temperatures of crystals having very different sizes but with the same structure. 
This is a possible interpretation of the phenomenon. It is rather difficult to deny this scenario in a clear manner 
since nano-crystals are formed in a supercooed metastable liquid state anyway.  This is the source of the long-lasting serious controversies on the nature of the glacial phase 
\cite{tanaka2013importance}. 
In the main text, we describe a few reasons why this scenario is difficult to explain our observation. 
Here we further discuss this possibility and show convincing experimental evidence that this transition cannot be explained by nano-crystal formation 
but should be LLT. 

If we try to explain the endothermic peak by the nano-crystal scenario, we need to assign this peak as (i) the melting of very small crystals, (ii) the melting of a new type of crystal 
distinct from the known crystal, or (iii) the solid-state transition from an ordered to a disordered crystal. 
First of all, there is no sign of the glass transition associated with liquid 1 after the formation of the glacial phase (see Fig. 1b in the main text). 
In the nano-crystal scenario, the glacial phase is considered as a mixture of glass 1 and nano-crystals. Thus, the absence of the glass transition 
of liquid 1 means that the system is filled with nano-crystals without any amorphous parts. This is not consistent with a much smaller heat of fusion of the glacial phase than crystals (see Fig. 1b in the main text) 
and previous X-ray scattering experimental results \cite{hedoux1999mesoscopic,derollez2004structural,mei2004local,murata_xray}. 
However, one may still argue that crystals are so disordered and thus the heat of fusion is much smaller than that of good crystals (see below). 
The melting of such small unstable nano-crystals might happen at a much lower temperature than $T_m$ \cite{hedoux2002conversion}. 
Then the endothermic peak might be the melting of such extremely small defective crystals. 
In this scenario, however, the gradual change of the glass transition temperature as a function of $T_{\rm rc}$ cannot be explained 
since there should be only liquid 1 that contributes to the glass transition.
Furthermore, according to the X-ray scattering measurements, there is any indication of neither the formation of a new type of crystals, 
nor the disappearance or position change of the Bragg peaks around 250 K. 
All these indicate that the above-mentioned scenarios based on nano-crystals cannot explain our DSC results. 

Next we show the annealing temperature $T_a$-dependence of the heating curve of TPP samples 
after annealed for the same fixed duration of 600 min, in 
Supplementary Figure~\ref{fig:tw600min}. 
Note that the transition is completed above $T_a$ = 216 K before 600 min, whereas it is not completed below 216 K. 
Here we focus on the behaviours of the second transition appearing after the glass transition 
and the third transition around 300 K, which is the melting of crystals.   
We can see that the endothermic peak position of the second transition shifts towards a higher temperature 
with an increase in $T_a$. The peak eventually disappears for $T_{\rm a}>232$ K and the highest observable peak temperature 
is located around 270 K. 
On the other hand, the melting of bulk crystals always takes place around 300 K irrespective of $T_a$.

We summarize the heat associated with phase transitions (the reverse LLT, crystallization during heating, and the crystal melting) in 
Supplementary Figure \ref{fig:tw600result}a 
and the peak temperatures of the reverse LLT and the crystal melting in 
Supplementary Figure \ref{fig:tw600result}b. 
The peak temperature of the endothermic peak of the reverse LLT monotonically increases with an increase in $T_{\rm a}$. 
Below $T_{\rm a} \sim 220$ K, the effect may be intrinsic, but above 220 K the peak temperature shift might be due to the presence of nano-crystals 
embedded in glass 2. The former is because there are few nano-crystals formed during annealing for $T_{\rm a} <220$ K. 
On the other hand, the latter is because the amount of nano-crystals formed during the heating, 
which can be estimated by subtracting the heat released upon crystallization from the heat absorbed upon melting (see Supplementary Figure 2a), starts to increase above 220~K
with an increase in $T_{\rm a}$  
The increase in the onset temperature of the glass transition of liquid 2 and the reverse LLT for $T_{\rm a}<220$ K may be due to the effects of ageing, since 
the ageing of glass 2 proceeds more quickly at a higher $T_{\rm a}$. 
In relation to this origin of the temperature shift, it should be noted that according to our previous X-ray scattering studies, the number density of locally favoured structures, 
or the order parameter $S$, does not depend on the annealing temperature $T_{\rm a}$.  
Since the $T_{\rm a}$-dependence of the endothermic position is not fully understood, however, we need further investigation to clarify its origin.  

If we assume that the endothermic peak is due to the nano-crystal melting, 
the shift of the peak position towards a higher temperature can be explained by the larger size of crystals and/or their higher perfectness for higher $T_a$. 
This scenario suggests that the melting peak of nano-crystals should continuously shift towards that of bulk crystal. 
However, our data shows that the endothermic peak cannot exist above 270 K, indicating its discontinuous jump 
to the melting peak of bulk crystal between $T_{\rm a}=232$ K 
and 237 K (see Supplementary Figure 1). 
This discontinuity between the endothermic peak and the melting peak of 
bulk crystals suggests that the endothermic peak is not due to nano-crystal melting.

We also note that the fact that the endothermic peak due to the reverse LLT disappears above 232 K indicates that 
only below this temperature LLT takes place. Thus, the binodal temperature of LLT is determined as $T_{\rm BN} \sim$232 K. 
This value of $T_{\rm BN}$ is consistent with our previous estimation, $T_{\rm BN} \sim 230$ K  \cite{tanaka2004liquid,kurita2004critical}. 
If we anneal a sample above this temperature, we have only crystallization phenomena and LLT cannot be induced.

The results shown in 
Supplementary Figure~\ref{fig:tw600min} also show clearly that the onset temperature of the reverse LLT 
increases with an increase in $T_{\rm a}$, thus suggesting the increase of $T_{\rm g}^2$ with $T_{\rm a}$. 
We confirm that this is indeed the case. 
Supplementary Figure \ref{fig:Trc}a shows the $T_{\rm rc}$-dependence of the endothermic signal coming from  
the reverse LLT for $T_{\rm a}$=225~K. 
Supplementary Figure \ref{fig:Trc}b plots the $T_{rc}$-dependence of the heat released upon heating estimated from the results in Supplementary Figure \ref{fig:Trc}a. 
From this, we can estimate the onset of the reverse LLT to be located around 242~K, suggesting that $T_{\rm SD}^{2 \rightarrow 1} \sim$ $T_{\rm g}^2 \sim$ 242~K 
for glass 2 formed at $T_{\rm a}$=225 K. 
In the main text we show that $T_{\rm SD}^{2 \rightarrow 1} \sim$ 235~K for $T_{\rm a} \sim 216$ K. 
Thus, this result clearly indicates that $T_{\rm g}^2$ and $T_{\rm SD}^{2 \rightarrow 1}$ increase with an increase in $T_{\rm a}$. 

Next we show in Supplementary Figure~\ref{fig:225K} the time dependence of the DSC heating curve for TPP samples 
annealed at 225 K. At this temperature, the melting peak of bulk crystals around 300 K is very large compared to 
those for lower $T_{\rm a}$ and the amount of the crystals increases with an increase in the annealing time. 
On the other hand, the exothermic heat between the endothermic peak and the melting peak upon heating 
does not increase so much by increasing the annealing time, clearly indicating 
that the crystals which melt around 300 K is formed during LLT and not during heating.
However, it is not reasonable to assume that nano-crystals and bulk crystals are formed 
simultaneously in the annealing process.  
Nano-crystals can be formed only under the situation that crystals cannot grow after nucleation.  
Extremely low mobility or internal frustration makes such a situation possible, but should make the formation of bulk crystals impossible. 
Thus, we conclude that it is difficult to have nano-crystals and bulk crystals at the same time in an ordinary situation.  

Finally, we evaluate the transformation heat of each process: 
the endothermic peak due to the reverse LLT, the exothermic broad part due to crystallization, and the crystal melting peak 
at several annealing temperatures 
(see Supplementary Figure~\ref{fig:deltaE}). 
The time evolution of the endothermic heat reflects the process of the endothermic transition. 
It should be noted that the melting component grows before the endothermic peak grows. 
This trend is evident particularly at $T_{\rm a}$ below 216 K. 
For example, at $T_{\rm a}=210$ K, the melting component appears around the annealing time of 30-60 min, whereas the endothermic peak 
appears around 1000 min. 
This fact strongly suggests that the endothermic transition and crystallization are the processes of essentially different nature.  
This also indicates the difficulty of the nano-crystal scenario and supports the LLT scenario.

In relation to the above, it should be noted that the crystal formation before the transition has been observed 
by neither light scattering nor X-ray scattering measurements. 
We show in Supplementary Figure~\ref{fig:raman} Raman scattering data at various $T_a$'s, which are taken after LLT is completed. 
There are small peaks due to crystals in the data above $T_a$ = 225 K,  
whereas such peaks are absent below 223 K and the spectra are typical amorphous signals, 
indicating the absence of crystals. 
These results indicate that crystals detected by our DSC in the process of the transformation during annealing 
should be unusually small, i.e., nano-crystals.

\vspace*{1cm}
\noindent
{\bf Supplemetary Note 2: 
Nucleation-growth and spinodal-decomposition-type LLT revealed by the glass-transition behaviours}

According to our optical microscopy observation \cite{tanaka2004liquid}, 
there are two types of the dynamic processes in LLT of TPP: nucleation-growth (NG)-type and spinodal-decomposition (SD)-type 
LLT. This classification is based on the types of pattern evolution. 
We confirm the presence of these two types of LLT from the glass-transition behaviour of TPP during the 
LLT process, as discussed in the main text.  
We stress that the glass transition behaviour is specific to a liquid state. 
The existence of the two glass transition temperatures for a single-component liquid and the presence of two types (NG-type and SD-type)  
temporal changes of the glass transition temperatures strongly support that the transition is indeed LLT and the endothermic peak observed 
on heating is the reverse LLT. 


Each feature of NG-type and SD-type LLT appears in the behaviour of 
the glass transition temperature, $T_{\rm g}$. 
Supplementary Figure~\ref{fig:onsetTg1-all} is the annealing time dependence of 
the glass transition temperature for several annealing 
temperatures. We estimate $T_{\rm g}$ as the onset temperature of the glass transition upon 
heating. The method to estimate $T_{g}$ is shown in the inset of the panel of $T_{\rm a}=212$ K. 
The $T_{\rm g}$ determined in this way is shown by an open circle on each curve. 
We analyse only the data after ageing of liquid 1 to see the change of the glass transition temperature due to LLT alone. 
Above 216 K, $T_{g}$ shows almost no change with the progress of LLT, indicating that this glass transition is 
always that of liquid 1. The magnitude of the glass transition step, or the amount of liquid 1, monotonically decreases with the annealing time, 
reflecting the nucleation and growth of liquid 2 domains in the sample, which was observed with optical microscopy \cite{tanaka2004liquid}.  
This is the typical behaviour expected for the NG-type LLT.    
On the other hand, $T_{g}$ continuously shifts towards a higher temperature 
with the waiting time for $T_{\rm a}$ below 212 K, suggesting that the transition from liquid 1 to liquid 2 is continuous.  
This behaviour is consistent with SD-type LLT. 
These results clearly indicate that the dynamical process of LLT can be classified into NG-type and SD-type (see also Fig. 4 in the main text). 
This is fully consistent with our microscopic observation of pattern evolution \cite{tanaka2004liquid}, 
strongly supporting the LLT scenario \cite{tanaka2000general}. 

Supplementary Figure \ref{fig:reverse} shows 
the $T_{\rm rc}$-dependence of the glass transition behaviours during the reverse LLT. 
The gradual continuous change of $T_{\rm g}$ during the reverse LLT transition suggests that this process may be 
SD-type transformation, which is consistent with its rather rapid transformation.

Finally we show the effects of ageing on the onset temperature of the glass transition by taking the data of $T_{\rm a}=210$ K as an example. 
Supplementary Figure \ref{fig:ageing}a shows 
the DSC curves in the glass transition region for $T_{\rm a}=210$ K, 
whereas Supplementary Figure \ref{fig:ageing}b plots the 
estimated onset temperature against the annealing time $t_{\rm w}$. We can see that the initial increase of $T_{\rm g}$ is due to the ageing 
of glass 1. The time region of the constant $T_{\rm g}$ of glass 1 indicates that the ageing of glass 1 is completed after 10$^{-1}$ min. 
The final continuous increase after 10$^3$ min is due to SD-type LLT, which can be seen in Supplementary Figure \ref{fig:ageing}b.

%
\end{document}